\documentclass[pre,preprintnumbers,amsmath,amssymb,superscriptaddress,onecolumn]{revtex4}
\usepackage{graphicx}

\def\R{{\mathbb R}}

\def\P{{\mathbb P}}
\def\pa{{\partial\Omega}}
\def\l{\ell}

\def\r{{\bf r}}
\def\s{{\bf s}}

\def\ve{\varepsilon}

\def\erfc{{\rm erfc}}

\begin{document}

\title{Efficient Monte Carlo methods for simulating \\ diffusion-reaction processes in complex systems}

\author{Denis Grebenkov} 
 \email{denis.grebenkov@polytechnique.edu}
\affiliation{
Laboratoire de Physique de la Mati\`ere Condens\'ee, \\ CNRS -- Ecole Polytechnique, 91128 Palaiseau, France \\
denis.grebenkov@polytechnique.edu}

\begin{abstract}
We briefly review the principles, mathematical bases, numerical
shortcuts and applications of fast random walk (FRW) algorithms.  This
Monte Carlo technique allows one to simulate individual trajectories
of diffusing particles in order to study various probabilistic
characteristics (harmonic measure, first passage/exit time
distribution, reaction rates, search times and strategies, etc.) and
to solve the related partial differential equations.  The adaptive
character and flexibility of FRWs make them particularly efficient for
simulating diffusive processes in porous, multiscale, heterogeneous,
disordered or irregularly-shaped media.
\end{abstract}

\date{\today}

\maketitle

\section{Introduction}
\label{sec:intro}

Diffusion in complex systems often invokes numerical simulations.
Except few simple shapes (such as a disk or a sphere) for which
diffusion equations possess explicit solutions \cite{Crank,Carslaw},
one needs to resort to numerical methods that can be roughly divided
into two groups:
\begin{enumerate}
\item
{\bf Finite differences, finite elements, boundary elements, etc.}  A
domain and/or its boundary are discretized with a regular or adaptive
mesh.  The original continuous problem is then replaced by a set of
linear equations to be solved numerically.  The solution is obtained
at all mesh nodes at successive time moments.  Since the accuracy and
efficiency of these deterministic numerical schemes significantly rely
on the discretization, mesh construction turns out to be the key issue
and often a limiting factor, especially in three dimensions.

\item
{\bf Monte Carlo simulations}.  A probabilistic interpretation of
diffusion equations is employed
\cite{Ito,Freidlin,Bass,Feller,Redner,Hughes,Weiss} to represent the
original continuous problem as the expectation of a functional of an
appropriate stochastic process.  Many random trajectories of this
process are then generated and used to approximate the expectation and
thus the solution.  Since there is no discretization, neither of the
domain, nor of boundary conditions, Monte Carlo techniques are
flexible and easy to implement, especially for studying diffusion in
complex geometries \cite{Sabelfeld,Sabelfeld2,Milshtein}.
\end{enumerate}

In basic Monte Carlo simulations, one fixes a time step $\delta$ and
approximates each trajectory $\r(t)$ by a sequence of $t/\delta$
independent normally distributed random jumps along each coordinate,
with mean zero and variance $2D\delta$.  Note that other jump
distributions may be used, e.g., discrete displacements to neighboring
sites of a lattice, ballistic displacements in random direction, etc.
At each jump, one has to check whether the trajectory remains inside
the confining domain.  If not, the jump has to be modified according
to the imposed boundary condition.  For instance, Neumann boundary
condition is implemented by reflecting the trajectory into the domain,
while the simulation is stopped for Dirichlet boundary condition.
Partial absorption/reflections, trapping, splitting and other local
mechanisms can also be implemented.  These fixed-time step simulations
are easy to implement but are inefficient in hierarchical or {\it
multiscale} porous media.  In fact, the time step $\delta$ must be
chosen so small to ensure that the average one-step displacement
$\sqrt{2D\delta}$ is much smaller than the smallest geometrical
feature of the medium.  Since most particles are released inside large
pores, a very large number of steps, $t/\delta$, may be required for
simulating each trajectory.  This is the major drawback of basic Monte
Carlo schemes.

\begin{figure}
\begin{center}
\includegraphics[width=80mm]{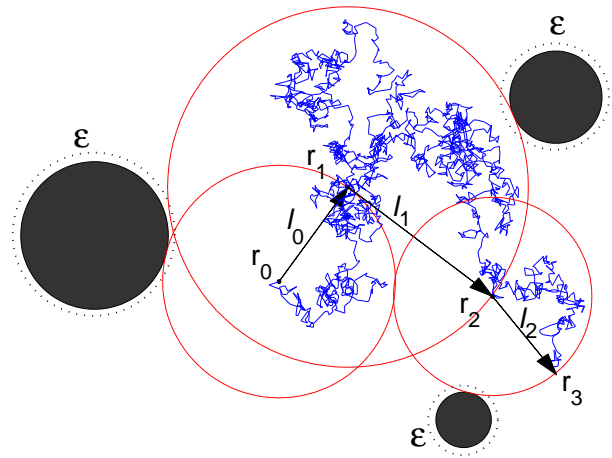}
\end{center}
\caption{
A fast random walk in a medium with obstacles (dark disks).  From an
initial position $\r_0$, one determines the distance $\ell_0$ to the
obstacles (or to the boundary of the medium) and draws a circle of
radius $\ell_0$ centered at $\r_0$.  A ``jump'' to a randomly
(uniformly) chosen point on the circle is then executed.  This single
large displacement (shown by an arrow) replaces a detailed simulation
of Brownian trajectory inside the disk of radius $\ell_0$.  From the
new point $\r_1$, one determines the distance again and executes the
next jump, and so on (only three jumps are shown) \cite{Grebenkov11}.
}
\label{fig:fast}
\end{figure}

In order to overcome this limitation, M\"uller proposed the concept of
variable time-step, geometry-adapted or fast random walks (FRWs)
\cite{Muller56}.  This technique was broadly employed by many authors,
for instance, to simulate diffusion-limited growth phenomena
\cite{Witten81,Ossadnik91,Halsey00} and to study diffusion-reaction processes
and related first-passage problems in random packs of spheres
\cite{Torquato89,Zheng89,Lee89} or near prefractal boundaries
\cite{Grebenkov05a,Grebenkov05b,Grebenkov06c,Levitz06}.  The idea is
to replace Brownian motion by an equivalent ``spherical process'' that
explores the medium as fast as possible.  A particle starts to diffuse
from an initial position $\r_0$ which may be prescribed (fixed) or
chosen randomly.  One draws the largest disk (or ball in three
dimensions) which is centered at $\r_0$ and inscribed in the confining
medium (Fig.~\ref{fig:fast}).  Its radius $\l_0$ is the distance
between $\r_0$ and the boundary.  After wandering inside the disk
during a random time $\tau_1$, the particle exits the disk at random
point $\r_1$.  Since there was no ``obstacles'' inside the disk, all
the exit points of the disk are equally accessible for isotropic
Brownian motion so that the exit point $\r_1$ has a uniform
distribution on the circle of radius $\l_0$.  From $\r_1$, the new
largest disk of radius $\l_1$ is inscribed in the medium.  After
wandering inside the disk during a random time $\tau_2$, the particle
exits at random point $\r_2$, and so on.  Following the Brownian
trajectory of the particle, one can construct the sequence of
inscribed disks (i.e., their centers $\r_n$ and radii $\l_n$) and the
associated exit times $\tau_n$.

The fundamental idea behind FRWs is that the sequence $\{ \r_n, \l_n,
\tau_n\}$ can be constructed directly, without simulating the underlying
Brownian trajectory at all.  At each step, one determines the distance
$\l_n$ between the current position $\r_n$ and the boundary of the
medium and chooses the next position $\r_{n+1}$ randomly and uniformly
on the circle of radius $\l_n$.  The time $\tau_{n+1}$ needed to exit
from the disk (i.e., to jump from $\r_n$ to $\r_{n+1}$) is a random
variable which can be generated from the well-known probability
distribution (Section \ref{sec:exit}).  A detailed time-consuming
simulation of a Brownian trajectory with high spatial resolution is
therefore replaced by generation of random jumps which are adapted to
the local geometrical structure of the medium.  In other words, the
spatial resolution of the spherical process is constantly adapted to
the distance to the boundary: closer the particle to the boundary,
finer the simulation scale.  Performing each jump at largest possible
distance yields a tremendous gain in computational time.

In this chapter, we review some practical aspects for an efficient
implementation of FRWs and discuss several extensions and applications
to first-passage phenomena.  Section \ref{sec:exit} introduces the
formulas for generating first exit times and positions.  In Section
\ref{sec:distance}, several strategies for estimating the distance to
the boundary are discussed.  Finally, Section \ref{sec:extensions}
presents extensions, applications and conclusions.

\section{Exit time and position distributions}
\label{sec:exit}

We start by recalling the computation of the first exit time and
position distributions for a general bounded domain $\Omega\subset
\R^d$.  Since the derived distributions will be used to generate jumps
in a FRW algorithm, the underlying ``jump'' domain $\Omega$ has to be
simple (e.g., to be a disk or a sphere, as described in Section
\ref{sec:intro}).  We emphasize that the simple shape of $\Omega$ and
the specific condition on its boundary $\pa$ have nothing to do with
the shape of the medium and the boundary mechanism that are going to
be modeled with the FRW algorithm.

We introduce the diffusion propagator $G_t(\r_0,\r)$ (also known as
heat kernel, or Green function of diffusion equation) as the
probability density for Brownian motion to move from a point $\r_0$ to
a vicinity of a point $\r$ in time $t$, {\it without hitting the
boundary $\pa$ of the domain $\Omega$ during this time}.  The
diffusion propagator satisfies the diffusion equation
\begin{equation}
\label{eq:diffusion}
\frac{\partial}{\partial t} G_t(\r_0,\r) = D \Delta G_t(\r_0,\r)  ,
\end{equation}
where $\Delta = \partial^2/\partial x_1^2 + ... + \partial^2/\partial
x_d^2$ is the Laplace operator, and $D$ the diffusion coefficient.
This equation is completed by the initial condition with Dirac
$\delta$-distribution, $G_{t=0}(\r_0,\r) = \delta(\r_0-\r)$, stating
that $\r_0$ is the starting point, and by Dirichlet boundary
condition, $G_t(\r_0,\r) = 0$ at $\r \in \pa$, that ``excludes''
Brownian trajectories that hit or crossed the boundary $\pa$ prior to
time $t$.  The diffusion propagator can be expressed in terms of the
($L_2$-normalized) Laplace operator eigenfunctions $u_m(\r)$ and
eigenvalues $\lambda_m$ \cite{Crank,Carslaw,Grebenkov13}
\begin{equation}
\label{eq:spectral}
G_t(\r_0,\r) = \sum\limits_{m=1}^\infty e^{-D\lambda_m t} u_m(\r_0) u_m^*(\r) ,
\end{equation}
where asterisk denotes the complex conjugate, and the eigenvalue
equation $\Delta u_m(\r) + \lambda_m u_m(\r) = 0$ is completed by
Dirichlet boundary condition $u_m(\r) = 0$ at $\r\in\pa$.

The diffusion propagator allows one to compute the first exit time and
position distributions.  For Brownian motion started from a point
$\r_0$, we denote $\tau = \inf\{ t>0~:~ \r(t)\notin \Omega\}$ the
first exit time $\tau$ from the domain $\Omega$.  Since $\tau$ is the
{\it first} exit time, Brownian motion should not hit the boundary
$\pa$ at earlier times $t < \tau$.  The probability of this event
(so-called survival probability) is obtained by integrating the
probability density $G_t(\r_0,\r)$ over all the arrival points $\r$:
\begin{equation}
\label{eq:St}
\P_{\r_0}\{ \tau > t\} \equiv S_{\r_0}(t) = \int\limits_\Omega d\r ~ G_t(\r_0,\r) .
\end{equation}
The probability density is simply
\begin{equation}
\label{eq:rhot}
\rho_{\r_0}(t) \equiv - \frac{\partial}{\partial t} S_{\r_0}(t) = 
- \int\limits_\Omega d\r ~ D\Delta G_t(\r_0,\r) =
\int\limits_\pa d\s ~ \left(-D\frac{\partial}{\partial n} G_t(\r_0,\r)\right)_{\r=\s} ,
\end{equation}
where we first used the diffusion equation and then Green formula to
integrate by parts.  Here $\partial/\partial n$ is the normal
derivative directed outwards the domain $\Omega$.  The probability
density $\rho_{\r_0}(t)$ characterizes the time of the exit,
regardless its position.  Alternatively, one can consider the exit
position $\s\in\pa$, regardless its time, whose distribution is known
as the harmonic measure \cite{Garnett}.  The harmonic measure density,
$\rho_{\r_0}(\s)$, can be expressed through the diffusion propagator
as
\begin{equation}
\label{eq:rhos}
\rho_{\r_0}(\s) = \int\limits_0^\infty dt  \left(-D\frac{\partial}{\partial n}  G_t(\r_0,\r) \right)_{\r = \s} .
\end{equation}
Looking at Eqs. (\ref{eq:rhot}, \ref{eq:rhos}), one can see that the
diffusive flux $-D \frac{\partial}{\partial n} G_t(\r_0,\r)$ plays the
role of the joint probability density for the exit time and position,
while $\rho_{\r_0}(t)$ and $\rho_{\r_0}(\s)$ are marginal densities
for time and position obtained after integration over the other
variable.  When both the exit time and position are needed, one can
use the joint distribution.  In practice, one can first generate the
exit time and then generate the exit position conditioned to be at the
prescribed exit time $t$:
\begin{equation}
\label{eq:rho_st}
\rho_{\r_0,t}(\s) = \frac{1}{\rho_{\r_0}(t)} \left(-D \frac{\partial}{\partial n} G_t(\r_0,\r) \right)_{\r = \s} ,
\end{equation}
where the prefactor $1/\rho_{\r_0}(t)$ reflects the conditional
character of this distribution.  Finally, one may also need the
probability density of the arrival positions $\r$ inside the domain
conditioned to survive up to time $t$, namely,
$G_t(\r_0,\r)/S_{\r_0}(t)$.

The above probability densities can be analytically computed for
simple domains for which the Laplace operator eigenfunctions are known
explicitly.  In the following three subsections, we summarize the
formulas for interval, disk and sphere.  For convenience, we will use
the rescaled (dimensionless) time and space variables, $t\to Dt/L^2$
and $x\to xL$, where $L$ is the length of the interval, or the radius
of the disk/sphere.

\subsection{Interval}

For the unit interval $(0,1)$, the eigenfunctions are eigenvalues of
the Laplace operator with Dirichlet boundary condition are
\begin{equation}
u_m(x) = \sqrt{2}~ \sin(\pi m x),  \qquad  \lambda_m = \pi^2 m^2 \quad (m = 1,2,3,\ldots),
\end{equation}
so that the one-dimensional propagator is simply
\begin{equation}
\label{eq:Gtau1}
G_t(x_0,x) = 2\sum\limits_{m=1}^{\infty} \sin(\pi m x_0) ~\sin(\pi m x)~ e^{-\pi^2 m^2 t} .
\end{equation}
It is also useful to write an alternative representation by image
method:
\begin{equation}
\label{eq:Gtau2}
G_t(x_0,x) = \frac{1}{\sqrt{4\pi t}}  \sum\limits_{k=-\infty}^{\infty} 
\left[e^{-(x-x_0 + 2k)^2/(4t)} - e^{-(x+x_0 + 2k)^2/(4t)}\right].
\end{equation}
The first expansion (\ref{eq:Gtau1}) rapidly converges for large $t$,
while Eq. (\ref{eq:Gtau2}) rapidly converges for small $t$.

Using Eq. (\ref{eq:St}), one gets two alternative representations for
the cumulative distribution function for the exit time:
\begin{equation}
\label{eq:S1d}
\begin{split}
S_{x_0}(t) & =
2\sum\limits_{m=1}^{\infty} \sin(\pi m x_0) \frac{1 -(-1)^m}{\pi m} ~e^{-\pi^2 m^2 t} \\
& = 1 - \erfc\left(\frac{x_0}{\sqrt{4t}}\right) + \sum\limits_{k=1}^\infty (-1)^k \left[\erfc\left(\frac{k - x_0}{\sqrt{4t}}\right)
- \erfc\left(\frac{k + x_0}{\sqrt{4t}}\right)\right], \\
\end{split}
\end{equation}
where $\erfc(x)$ is the complementary error function.  The density is
\begin{equation}
\begin{split}
\rho_{x_0}(t) & =  2\pi \sum\limits_{m=1}^{\infty} \sin(\pi m x_0) (1 -(-1)^m) m ~e^{-\pi^2 m^2 t}  \\
& = \frac{1}{\sqrt{4\pi t^3}} \sum\limits_{k=-\infty}^\infty (-1)^k (x_0 + k) \exp\left(-\frac{(x_0+k)^2}{4t}\right).  \\
\end{split}
\end{equation}
As the boundary of an interval consists of two endpoints, the harmonic
measure density is reduced to two functions, namely, the probability
$\rho_{x_0}(\s=0)$ to reach the left endpoint before the right one,
and that of the complementary event:
\begin{equation}
\rho_{x_0}(\s=0) = 2\sum\limits_{m=1}^\infty \frac{\sin(\pi m x_0)}{\pi m} = 1 - x_0 , \qquad  \rho_{x_0}(\s=1) = x_0 .
\end{equation}
This is the classical result for the gambler's ruin \cite{Feller}.
Other probability densities can also be explicitly written.

\subsection{Disk}

For the unit disk with Dirichlet boundary condition, the Laplace
operator eigenvalues and eigenfunctions are \cite{Crank,Carslaw}
\begin{equation}
\label{eq:u_2d}
\lambda_{nk} = \alpha_{nk}^2,  \qquad
u_{nk}(r,\theta) = \frac{\epsilon_n}{\sqrt{\pi}} ~ \frac{1}{-J_n'(\alpha_{nk})} ~ J_n(\alpha_{nk}r) \cos n\theta ,
\end{equation}
where $\epsilon_n = \sqrt{2}$ for $n>0$ and $\epsilon_0 = 1$,
$J_n'(z)$ is the derivative of the Bessel function $J_n(z)$ of the
first kind, and $\{\alpha_{nk}\}_{k=0,1,2,\ldots}$ is the set of all
positive zeros of the function $J_n(z)$ (with $n=0,1,2,\ldots$).  For
convenience, the double index $nk$ is used instead of the single index
$m$ to enumerate the eigenfunctions and eigenvalues.  The asymptotic
behavior of zeros $\alpha_{nk}$ is well known while their numerical
computation is straightforward by bisection method or Newton's method.
After this preliminary step, the spectral decomposition
(\ref{eq:spectral}) yields the diffusion propagator $G_t(\r_0,\r)$ and
the consequent probability densities: $\rho_{\r_0}(t)$,
$\rho_{\r_0}(\s)$ and $\rho_{\r_0,t}(\s)$.  Since the starting point
$\r_0$ is located in the center of the disk, only the terms with $n =
0$ (invariant under rotation) do contribute.  In particular, one gets
\begin{equation}
\label{eq:S0_2d}
S_c(t) = 2 \sum\limits_{k=0}^\infty \frac{e^{-\alpha_{0k}^2 t}}{\alpha_{0k} J_1(\alpha_{0k})} ,
\end{equation}
while the harmonic measure density is uniform: $\rho_c(\s) =
\frac{1}{2\pi}$, as all the boundary points are equivalent from the
center (here and throughout the text, the subscript 'c' refers to the
starting point at the center of the jump domain).

\subsection{Sphere}

The Laplace operator eigenvalues and eigenfunctions for the unit
sphere with Dirichlet boundary condition are \cite{Crank,Carslaw}
\begin{equation}
\label{eq:u_3d}
\lambda_{nk} = \alpha_{nk}^2,  \qquad
u_{nkl}(r,\theta,\varphi) = \frac{\sqrt{2n+1}}{\sqrt{2\pi}} ~ \frac{1}{-j'_n(\alpha_{nk})} ~ j_n(\alpha_{nk}r) P_n^{l}(\cos \theta) e^{il\varphi},
\end{equation}
where $j'_n(z)$ is the derivative of the spherical Bessel function
$j_n(z)$ of the first kind, $P_n^l(z)$ the associated Legendre
polynomial, and $\{\alpha_{nk}\}_{k=0,1,2,\ldots}$ the set of all
positive zeros of the function $j_n(z)$ (with $n=0,1,2,\ldots$).  The
last index $l$ ranges from $-n$ to $n$.  The spectral decomposition
(\ref{eq:spectral}) allows one to compute all the necessary
distributions.  When the starting point $\r_0$ is located at the
center, the formulas are simplified, e.g.,
\begin{equation}
\label{eq:S0_3d}
S_c(t) = 2 \sum\limits_{k=1}^\infty (-1)^{k+1} e^{- \pi^2 k^2 t}
\end{equation}
(as $\alpha_{0k} = \pi(k+1)$, $k = 0,1,2,\ldots$), while the harmonic
measure density is uniform: $\rho_c(\s) = \frac{1}{4\pi}$.  Using the
techniques of Laplace transforms and series summation, one can get an
alternative representation for the survival probability as
\cite{Grebenkov07d}
\begin{equation}
\label{eq:S0_3da}
S_c(t) = 1 - \frac{2}{\sqrt{\pi t}} \sum\limits_{k=1}^{\infty} e^{-(2k-1)^2/(4t)} ,
\end{equation}
which rapidly converges for small $t$.

\subsection{Behavior of the exit time distribution}

\begin{figure}
\begin{center}
\includegraphics[width=80mm]{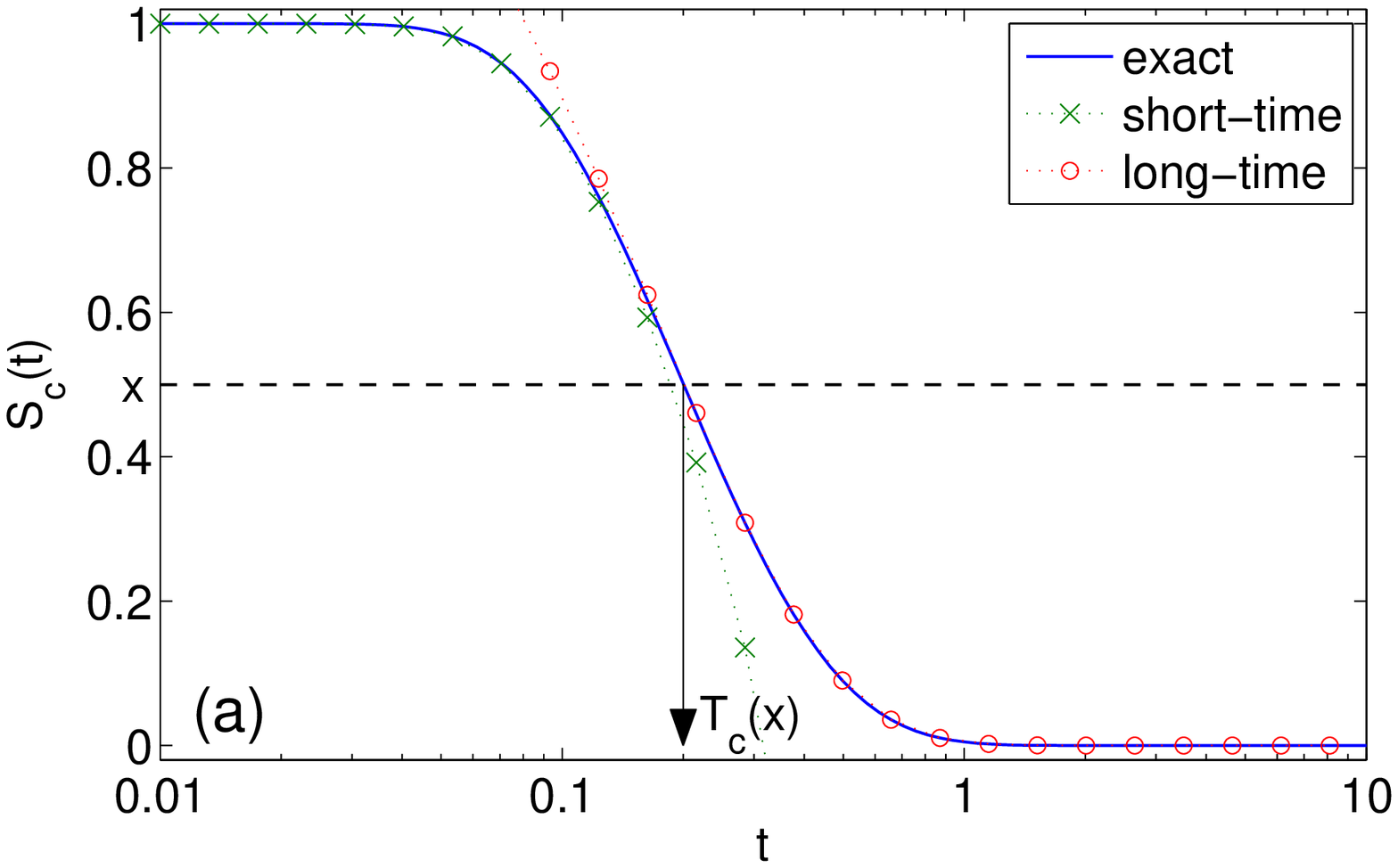}
\includegraphics[width=80mm]{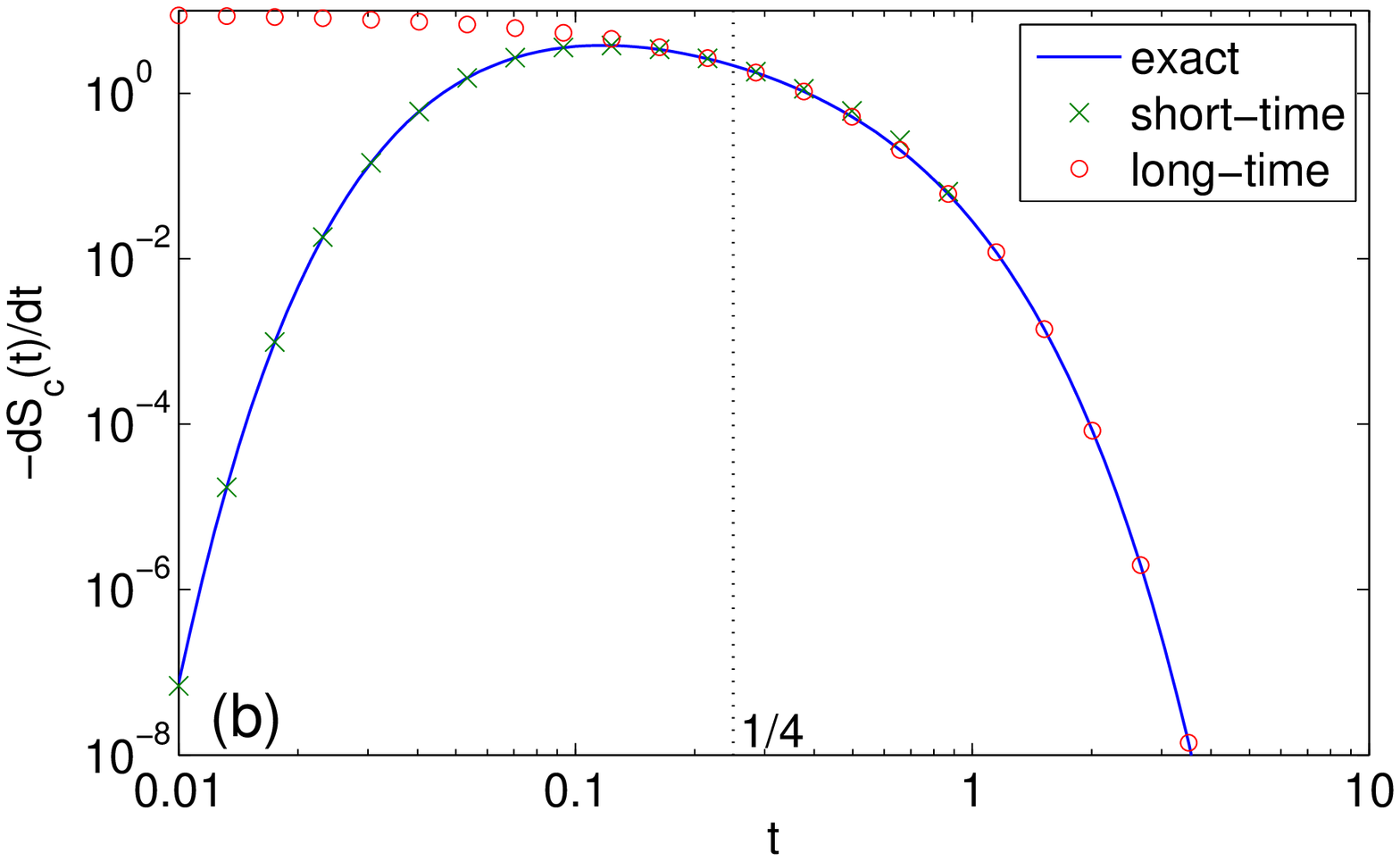}
\end{center}
\caption{
{\bf (a)}: The survival probability $S_c(t)$ and its short-time and
long-time asymptotic behaviors (\ref{eq:S0_short}, \ref{eq:S0_long})
for the unit disk.  The dashed horizontal line at $S_c(t) = x$
illustrates the construction of the inverse function $T_c(x)$.  {\bf
(b)}: The probability density $-dS_c(t)/dt$ of the exit time and its
short-time and long-time asymptotic behaviors derived from
Eqs. (\ref{eq:S0_short}, \ref{eq:S0_long}).  The vertical dotted line
shows the mean exit time $1/4$ \cite{Grebenkov11}. }
\label{fig:S0}
\end{figure}

By definition, the survival probability monotonously decreases from
$1$ at $t = 0$ to $0$ as $t$ goes to infinity, as illustrated on
Fig. \ref{fig:S0}a.  In the short-time limit, the exit probability $1
- S_c(t)$ is extremely small,
\begin{equation}
\label{eq:S0_short}
1 - S_c(t) \simeq \begin{cases}
2 e^{-1/(16t)} \frac{4\sqrt{t}}{\sqrt{\pi}}\bigl(1 - 8t + 192 t^2 + O(t^3)\bigr)  \hskip 6mm  (d=1) , \\
2 e^{-1/(4t)} \bigl(1 - t + 4t^2 + O(t^3)\bigr) \hskip 18mm (d=2), \\
2 e^{-1/(4t)} \frac{1}{\sqrt{\pi t} }  \hskip 44.5mm (d=3),  \\
\end{cases}
\end{equation}
because very few particles can travel the distance from the origin to
the boundary during a short time%
\footnote{
The prefactor is different for $d=1$ as the diameter of the unit
interval is twice smaller than that of the unit disk and sphere.  Note
also that, according to Eq. (\ref{eq:S0_3da}), the first correction
term for the case $d=3$ vanishes as $\exp(-9/(4t))$, i.e., much faster
than polynomial corrections in other dimensions.}.
In turn, for large $t$, the survival probability decays exponentially,
\begin{equation}
\label{eq:S0_long}
S_c(t) \simeq \begin{cases}
\frac{4}{\pi} e^{-\pi^2 t}  \hskip 17mm (d = 1) , \\
\frac{2}{\alpha_{00} J_1(\alpha_{00})} e^{-\alpha_{00}^2 t}  \hskip 3.5mm (d=2), \\
2e^{-\pi^2 t}  \hskip 18mm (d=3) , \\  \end{cases}
\end{equation}
since it is unlikely for diffusing particles to avoid the encounter
with the boundary during a long time.

Figure \ref{fig:S0}b shows the probability density, $\rho_c(t) =
-\frac{dS_c(t)}{dt}$, of the exit time.  This figure and the above
asymptotic behaviors clearly indicate that the exit time is localized
around its mean value which is equal to $1/(2d)$ (and $1/8$ in 1D).
In particular, the probability that $\tau$ does not belong to an
interval $(t_{\rm min}, t_{\rm max})$, can be made negligible by
choosing $t_{\rm min}$ and $t_{\rm max}$ appropriately.  For instance,
if $t_{\rm min} = 0.001$ and $t_{\rm max} = 10$, one has
\begin{equation}
\label{eq:Ptau}
\P\{ \tau \notin (0.001,10) \} = 1 - S_c(0.001) + S_c(10) < 10^{-12d}   \hskip 5mm (d = 2,3).
\end{equation}
As a consequence, the exit times beyond this interval can be ignored.

\subsection{Generation of the exit times}

The explicit form of Eqs. (\ref{eq:S0_2d}, \ref{eq:S0_3d}) allows one
to generate exit times by inversion method.  Inverting numerically the
function $S_c(t)$ (i.e., finding a function $T_c(x)$ such that
$S_c(T_c(x)) = x$ for any $x$ between $0$ and $1$), one obtains a
mapping from random variables $\chi_n$ with a uniform distribution on
the unit interval, to the exit times
\begin{equation}
\label{eq:tau}
\tau_n = \frac{\ell_n^2}{D} T_c(\chi_n) 
\end{equation}
for the interval of length $\ell_n$, or for the disk or sphere of
radius $\ell_n$, with a given diffusion coefficient $D$.  The uniform
random variables $\chi_n$ are generated by a standard routine for
pseudo-random numbers.

In summary, two preliminary numerical procedures are required for
generating the exit times:
\begin{enumerate}
\item
In 2D, one needs to find a finite number of positive zeros
$\{\alpha_{0k}\}$ of Bessel function $J_0(z)$.  The inequalities $\pi
k < \alpha_{0k} < \pi(k+1)$ allow one to search for a single zero on
each interval $(\pi k, \pi k + \pi)$ by bisection or Newton's method.
Since smaller times require larger truncation sizes, the asymptotic
formula (\ref{eq:S0_short}) can be used instead of the truncated
series in Eqs. (\ref{eq:S0_2d}, \ref{eq:S0_3d}) to improve the
accuracy at short times.  In one and three dimensions, this step is
skipped because the eigenvalues are known explicitly.

\item
One constructs the function $T_c(x)$ as a numerical solution $T_c(x) =
t$ of the equation $S_c(t) = x$ for a mesh of points $x$.  The
monotonous decrease of $S_c(t)$ ensures, for any $x$, the unique
solution which can be computed by bisection method or Newton's method.

\end{enumerate}
Both procedures rely on classical numerical methods.  Moreover, these
procedures have to be performed once and forever while the stored
values of the function $T_c(x)$ can be loaded before starting Monte
Carlo simulations.  As a consequence, the generation of the exit times
during simulations is reduced, through Eq. (\ref{eq:tau}), to a
routine generation of uniformly distributed pseudo-random numbers.

\subsection{Boundary condition}
\label{sec:boundary}

By construction, a generated sequence of disks or spheres may become
arbitrarily close to the boundary of the medium but it never hits the
boundary.  It is therefore necessary to set a threshold distance $\ve$
below which the trajectory is considered to hit the boundary
(Fig. \ref{fig:fast}).  This threshold is typically set to be much
smaller than any relevant length scale.  Since the average number of
jumps needed to reach the boundary scales as $\ln(1/\ve)$, setting
$\ve$ to very small values would not significantly slow down
simulations.

After the hit, the next step depends on the boundary condition or
mechanism to be modeled.  The two simplest conditions, Dirichlet and
Neumann, represent purely absorbing and purely reflecting boundaries,
respectively.  In the former case, once a particle arrives onto the
boundary, it is immediately ``killed'' and the simulation of the
trajectory is stopped.  This absorbing condition may model various
physical, chemical or biological mechanisms, e.g., relaxation of local
magnetization on paramagnetic impurities dispersed on the interface in
NMR experiments; one-way permeation through cellular membranes;
chemical transformation on the catalytic surface,
etc. \cite{Brownstein79,Bond,Berg81,Adam,Sano81}.  Whatever the
microscopic mechanism is, the interaction with the boundary changes
the state of the particle and thus removes it from the transport
process.  In the opposite case of Neumann boundary condition, the
particle is reflected back into the medium and continues its diffusive
motion, without changing its state.  One can also simulate the
so-called Robin boundary condition for partially absorbing/reflecting
boundary when absorption and reflection events are chosen randomly
\cite{Sapoval94,Grebenkov06,Grebenkov06c}.  Note that the boundary may
have variable properties in space and time, e.g., some parts of the
boundary may be reflecting while the other absorbing (this is the
typical situation for search problems when a target is located on the
boundary \cite{Singer06,Schuss07}); moreover, these properties may
evolve with time, e.g., during passivation/deactivation of catalysts
\cite{Filoche08}.  One can also consider multi-compartment media with
permeable boundary (e.g., living cells and the extracellular space).
Once the trajectory hits this boundary, it may be reflected on both
sides (with equal or different probabilities).  Apart from these
standard boundary mechanisms, one can model trapping of a particle for
a fixed or randomly distributed waiting time \cite{Holcman05},
multistage surface kinetics \cite{Chou07}, and surface transfer
mechanisms \cite{Benichou08,Benichou11}.  Finally, the FRW algorithm
for bulk diffusion can be combined with other numerical techniques for
modeling transport on the boundary.  Such a flexibility of FRW
algorithms makes them attractive for simulating various diffusive
phenomena.

\begin{figure}
\begin{center}
\includegraphics[width=90mm]{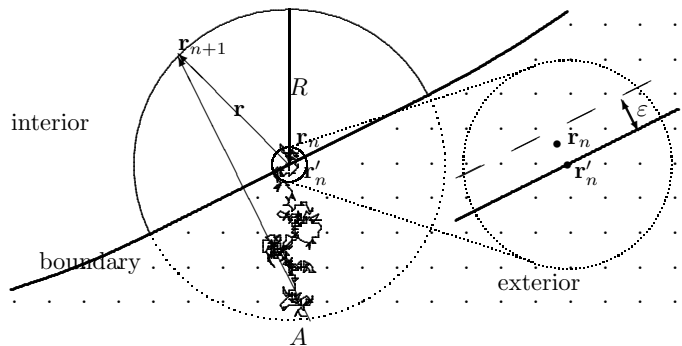}
\end{center}
\caption{
When the particle has approached the reflecting boundary of the medium
closer than a prescribed threshold $\ve$, it is ``released'' on a
circle of radius $R$ centered at the encounter boundary point $\r'_n$.
If the released point $A$ does not belong to the interior of the
medium, one uses its mirror reflected point $\r_{n+1}$ inside the
medium.  The radius $R$ should be chosen as large as possible, but
small enough in comparison to the characteristic length of the
boundary (so that the boundary is almost flat at scale $R$)
\cite{Grebenkov11}. }
\label{fig:demisphere}
\end{figure}

The implementation of Dirichlet boundary condition is trivial:
simulation of the trajectory is simply stopped after hitting the
boundary.  One can record the hitting position (for computing the
harmonic measure \cite{Grebenkov05a,Grebenkov05b}), the hitting time
(for computing the exit time distribution \cite{Levitz06}), the
traveled distance (for computing the time-dependent diffusion
coefficient \cite{Latour93,Latour94,deSwiet96,Sen04,Novikov11}), etc.

Neumann boundary condition is much more delicate, as the trajectory
has to be reflected back into the medium.  In the simplest
implementation, one just moves the current position to the interior
point of the medium at a fixed (small) distance from the boundary.
This reflection distance has to be larger than the hitting threshold
$\ve$ but much smaller than other relevant length scales.  The choice
of the reflection distance is a compromise between the accuracy and
rapidity of simulations: a large distance would bias the results,
while a small distance would lead to multiple small jumps after
reflection.  Another, much more efficient way to reflect back the
trajectory is to make a relatively large jump from the boundary of the
domain to a half-circle or half-sphere (Fig. \ref{fig:demisphere}).
Indeed, if the boundary is flat (or may be approximated as flat up to
a certain length scale), diffusion in the related half-disk
(half-ball) with the reflecting base is equivalent to diffusion in the
whole disk (ball).  For instance, the distribution of the exit times
is the same, while the distribution of the exit position is still
uniform.  An implementation of this reflection jump allows one to
significantly speed up simulations.

\subsection{Incomplete jump}

If the reflecting condition is set over the whole boundary, the
trajectory would never stop.  One adds therefore a condition to stop
the simulation when the time counter $t = \tau_1 + \tau_2 + \ldots +
\tau_n$ exceeds a prescribed time $T$.  More generally, whatever the
boundary condition is, one may need to stop the trajectory after or at
time $T$.  If the trajectory has to be stopped {\it precisely at} $T$
(e.g., to compute the traveled distance at time $T$), the last jump
has to be simulated differently.  In fact, we are interested in the
{\it intermediate} position $\r(T)$ of the trajectory which is started
at the center of the jump domain at time $t_0 = t - \tau_n < T$ and
conditioned to exit from this domain at time $t > T$.  The trajectory
can be split into two independent parts: from $t_0$ to $T$, and from
$T$ to $t$.  The conditional probability density for the intermediate
position $\r$ is
\begin{equation}
\frac{G_{T-t_0}(\r_0,\r) ~ \rho_{\r}(t-T)}{\rho_{\r_0}(t-t_0)} ,
\end{equation}
where the first factor is the probability density for moving from
$\r_0$ to $\r$ during the time $T-t_0$ (first part), the second factor
is the probability density for exiting the domain at time $t-T$ after
starting at $\r$ (second part), while the denominator is the
probability density for exiting at time $t-t_0$ after starting at
$\r_0$.  The intermediate position $\r$ can be generated from this
density.

\subsection{Extension: rectangular domains}

\begin{figure}
\begin{center}
\includegraphics[width=130mm]{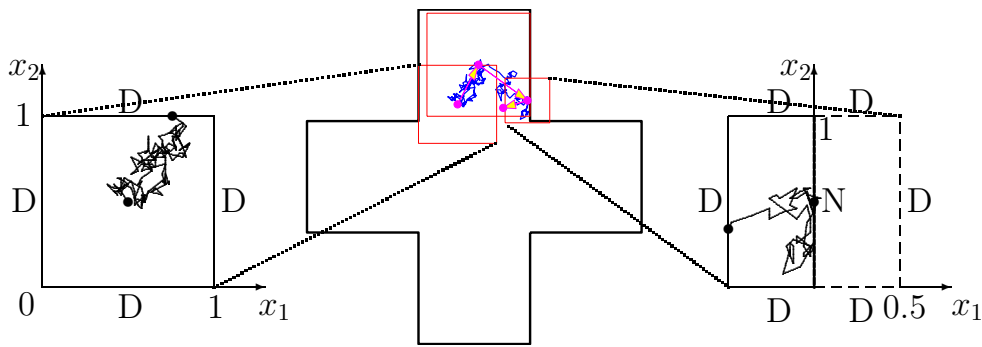}
\end{center}
\caption{
A FRW simulation of reflected Brownian motion inside a cross-shaped
medium by using square jump domains (shown in red) whose centers (full
circles) belong to the trajectory (shown by solid blue line and
generated by basic Monte Carlo simulations).  On the left, we show a
zoom of the first square jump domain for which the distributions of
the exit time and position can be explicitly found by solving the
appropriate PDE equation with Dirichlet boundary condition (denoted by
'D').  Each jump starts from the center.  On the right, we show the
third jump domain for reflection from the boundary.  The right
vertical edge contains the starting point and coincides with the
reflecting boundary of the medium.  The mirror reflection across this
vertical edge reduces the analysis to that on the square with
Dirichlet boundary condition. }
\label{fig:BM}
\end{figure}

The boundary of many model or image-reconstructed structures is formed
by horizontal and vertical segments (in 2D) or by rectangular plates
oriented along three coordinate axes (in 3D).  For such structures, it
may be more convenient to split the trajectory into ``elementary
blocks'' inside rectangular jump domains rather than spherical ones,
as illustrated on Fig. \ref{fig:BM} \cite{Deaconu06,Zein10}.  The use
of rectangles/parallelepipeds as jump domains allows one to
significantly reduce the number of small jumps near the boundary of
the medium.  On one hand, the generation of jumps becomes more
complicated because the arrival points over the edges are not
uniformly distributed any more.  One needs therefore to generate not
only the exit times but also the conditional exit positions according
to Eq. (\ref{eq:rho_st}).  On the other hand, Brownian motion inside
rectangles/parallelepipeds is formed by independent one-dimensional
Brownian motions along each coordinate, and the problem is essentially
reduced to one-dimensional setting on the interval.  In particular,
the diffusion propagator $G_t^{(d)}(\r_0,\r)$ in the jump domain
$\Omega = [0,L_1]\times [0,L_2]\times ... \times [0,L_d]\subset \R^d$
is factored out as
\begin{equation}
G_t^{(d)}(\r_0,\r) = \prod\limits_{j=1}^d \frac{1}{L_j} ~ G_{Dt/L_j^2}(\r_{0j}/L_j, \r_j/L_j) ,
\end{equation}
where $G_t(x_0,x)$ is the one-dimensional propagator in the unit
interval $[0,1]$ given by Eqs. (\ref{eq:Gtau1}, \ref{eq:Gtau2}), with
dimensionless time $Dt/L_j^2$ and dimensionless space coordinates
$x_0$ and $x$, representing the starting and arrival points.

The cumulative distribution of the exit time is the survival
probability in $\Omega$:
\begin{equation}
S_{\r_0}^{(d)}(t) = \int\limits_\Omega d\r ~ G_t^{(d)}(\r_0,\r) = \prod\limits_{j=1}^d S_{\r_{0j}/L_j}(Dt/L_j^2),
\end{equation}
where $S_{x_0}(t)$ is given by Eq. (\ref{eq:S1d}).  As usual, the
probability density is $\rho_{\r_0}^{(d)}(t) = -
\frac{\partial}{\partial t} S_{\r_0}^{(d)}(t)$.
In the case of the (hyper)cube (with $L_j = L$) with the starting
point in the center (i.e., $\r_0 = [L,L,...,L]/2$), one gets
\begin{equation}
\label{eq:Std}
S_{c}(t) = \bigl[S_{\frac12}(Dt/L^2)\bigr]^d,  \qquad   \rho_{c}(t) = \frac{Dd}{L^2} \bigl[S_{\frac12}(Dt/L^2)\bigr]^{d-1} \rho_{\frac12}(Dt/L^2).
\end{equation}

Once the exit time is generated, one needs to generate the exit
position at this time.  For the sake of clarity, we only consider the
starting point from the center of the domain.  Starting from
Eq. (\ref{eq:rho_st}) and skipping technical derivations, we obtain
the conditional probability density of the exit point $\s$ to lie on
the face which is orthogonal to the coordinate axis $j$:
\begin{equation}
\label{eq:omega}
\omega_{c,t}^j(\s) =  \frac{q_j(t)}{2} \prod\limits_{k\ne j}^d \frac{G_{Dt/L_k^2}(1/2,s_k/L_k)}{L_k ~ S_{\frac12}(Dt/L_k^2)} ,
\end{equation}
where the last factor is the product of conditional probability
densities for (independent) displacements in the orthogonal directions
$k\ne j$, and
\begin{equation}
\label{eq:qj}
q_j(t) \equiv \frac{L_j^{-2} ~\frac{\rho_{\frac12}(Dt/L_j^2)}{S_{\frac12}(Dt/L_j^2)}}{\sum\nolimits_{k=1}^d L_k^{-2} ~ 
\frac{\rho_{\frac12}(Dt/L_k^2)}{S_{\frac12}(Dt/L_k^2)}}
\end{equation}
is the conditional probability to exit through one of two faces at
direction $j$.  When all lengths are equal to each other, $L_1 = ... =
L_d = L$, one gets $q_j(t) = 1/d$ (all faces are equivalent from the
center).

The displacements along each coordinate can be generated by
numerically inverting the cumulative distribution
\begin{equation}
\label{eq:Fx}
\begin{split}
F_t(x) & \equiv \frac{1}{S_{\frac12}(t)} \int\limits_0^x dx' G_t(1/2,x') \\
& = \frac{2/\pi}{S_{\frac12}(t)} \sum\limits_{m=1}^\infty (-1)^{m+1} \frac{1-\cos(\pi(2m-1)x)}{2m-1} ~e^{-\pi^2(2m-1)^2 t}  \\
& =  \frac{1/2}{S_{\frac12}(t)} 
\sum\limits_{k=-\infty}^\infty \biggl[\erfc\left(\frac{x+2k+1/2}{\sqrt{4t}}\right)
- \erfc\left(\frac{2k+1/2}{\sqrt{4t}}\right) \\
& ~ \hspace*{22mm} - \erfc\left(\frac{x+2k-1/2}{\sqrt{4t}}\right) + \erfc\left(\frac{2k-1/2}{\sqrt{4t}}\right)\biggr]. \\
\end{split}
\end{equation}
A technical difficulty is that this function depends on two variables
$x$ and $t$, i.e., one needs to invert this function versus $x$ for
various values of $t$ as a parameter.  Other methods such as rejection
sampling or adaptive rejection sampling can also be applied to
generate these random variables.

As discussed in Section \ref{sec:boundary}, the boundary condition has
to be implemented.  When the particle is close to the reflecting
boundary, one needs to perform a reflection jump, as discussed in
Section \ref{sec:boundary}.  In this case, one or several edges
(faces) of the jump domain may be reflecting (Fig. \ref{fig:BM}).  In
order to generate the exit time and position, one can consider
diffusion in the extended jump domain which is composed of the
original jump domain and its mirror image with respect to the
reflecting boundary.  The previous formulas can be directly applied.

\section{Distance computation}
\label{sec:distance}

The adaptive splitting of the trajectory into independent ``elementary
blocks'' relies on estimating the distance from any interior point to
the boundary.  We emphasize that the jump distance may in practice be
smaller than the distance to the boundary, but it must not exceed this
distance.  In other words, the problem of computing the exact distance
can be relaxed to a simpler problem of finding a lower bound of the
distance.  Depending on the studied geometry, this purely geometrical
problem can be solved in different ways.  In this section, we briefly
describe two strategies for efficient distance estimation.

\subsection{Self-similar (fractal) domains}
\label{sec:GAFRW}

Fractals are often used as a paradigm of complex domains
\cite{Mandelbrot}.  On one hand, fractals are very irregular shapes
which exhibit geometrical details at various length scales,
``exploding'' the classical notions of length, surface area or volume.
On the other hand, self-similar or self-affine hierarchical structures
help to perform accurate theoretical and numerical analysis on
fractals.  In particular, the self-similarity of von Koch curves and
surfaces allows one for a rapid estimation of the distance from any
interior point to these boundaries \cite{Grebenkov05a,Grebenkov05b}.

\begin{figure}
\begin{center}
\includegraphics[width=110mm]{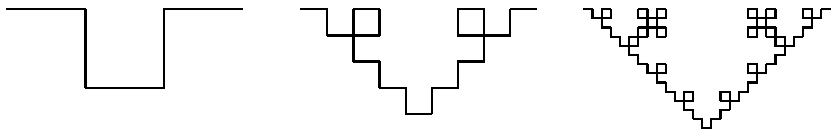}
\end{center}
\caption{
Three generations of the quadratic von Koch curve of fractal dimension
$\ln 5/\ln 3\approx 1.465$.  At each iteration, one replaces all
linear segments by the rescaled generator (first generation). }
\label{fig:koch_quad}
\end{figure}

\begin{figure}
\begin{center}
\includegraphics[width=100mm]{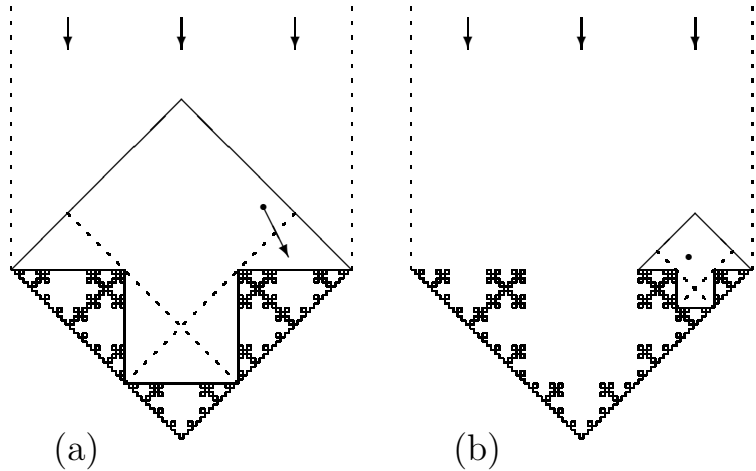}
\end{center}
\caption{
{\bf (a)} Basic arrow-like cell which is divided into the rotated
square and five small triangles.  Once a Brownian particle arrived
into the rotated square, the distance between its current position
(full circle) and the boundary of the arrow-like cell (the generator)
can be computed explicitly.  Random jumps inside the rotated square
can therefore be executed, until the particle either exits from the
arrow-like cell, or enters into a small triangle.  {\bf (b)} In the
latter case, the particle starts to ``see'' the geometrical details of
the next generation.  The rescaled arrow-like cell is then used to
compute the distance to the boundary.}
\label{fig:gafrw}
\end{figure}

To illustrate this idea, we consider the quadratic von Koch curve
(Fig.~\ref{fig:koch_quad}).  When a random walker is far from the
boundary, it does not ``distinguish'' its geometrical details.  One
can thus estimate the distance by considering the coarsest generation
(Fig.~\ref{fig:gafrw}).  Getting closer and closer to the boundary,
the random walker starts to ``recognize'' smaller and smaller
geometrical details.  But at the same time, when small details appear
in view, the rest of the boundary becomes ``invisible''.
Consequently, one can explicitly determine the distance by examining
only the local geometrical environment (see \cite{Grebenkov05b} for
details).  The advantage and eventual drawback of this
geometry-adapted algorithm is the need for a specific implementation
for each studied geometry.  Relying on self-similarity of von Koch
fractals, we were able to generate up to $10^{10}$ random trajectories
for highly irregular boundaries with up to 10-12 iterations (in 2D).
Note that these boundaries present geometrical features at length
scales from $1$ to $(1/3)^{10}$ (i.e., over five orders of magnitude).

\subsection{Sphere packs}

\begin{figure}
\begin{center}
\includegraphics[width=55mm]{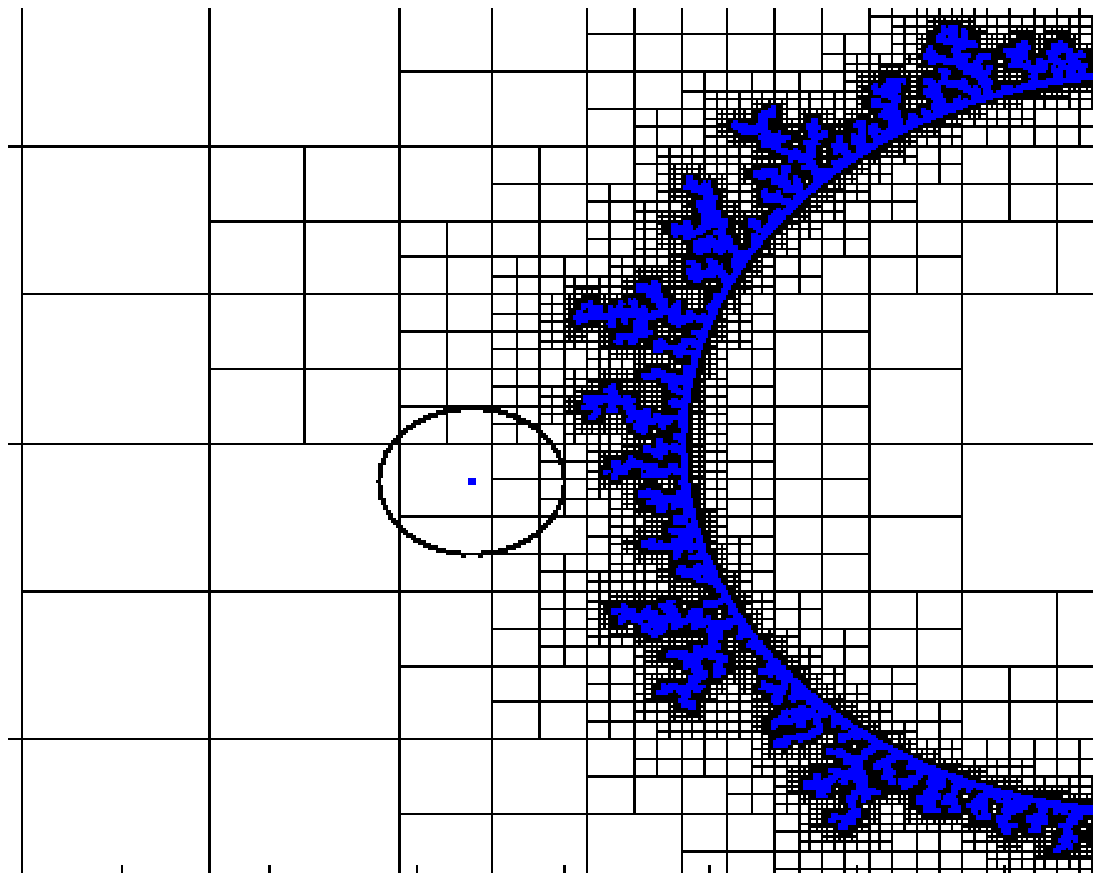}
\includegraphics[width=57mm]{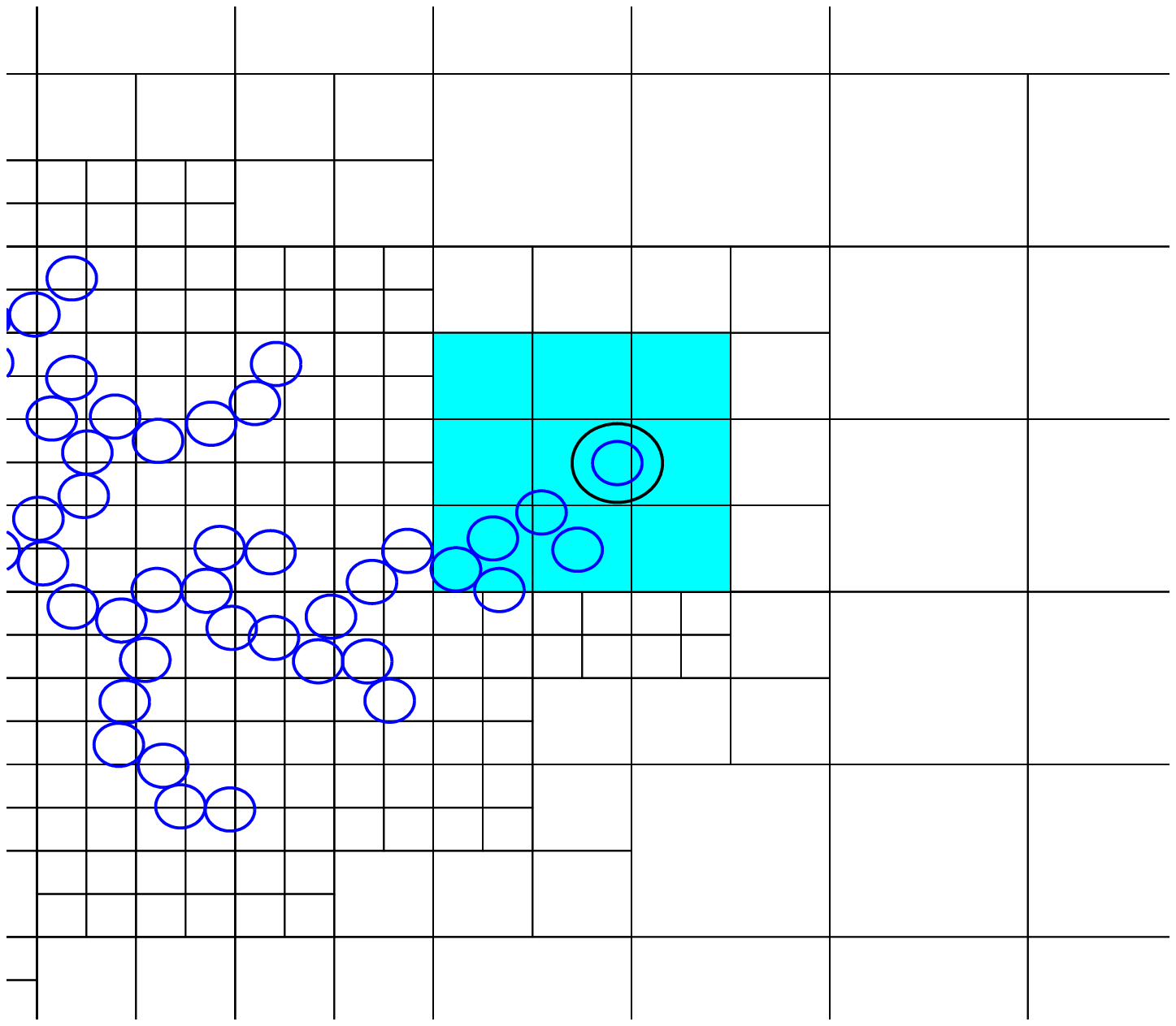}
\end{center}
\caption{
Whitney decomposition into dyadic squares near a DLA aggregate (shown
in blue), with a zoom on the right.  The finest resolution is set to
the diameter of small disks forming the aggregate (courtesy by
D. D. Nguyen).}
\label{fig:DLA_Whitney}
\end{figure}

Sphere packs are often used to model porous and granular media,
colloidal solutions, gels and polymer networks.  This is indeed a
quite generic model as spheres (or disks in two dimensions) can be
mono- or polydisperse, overlapping or not, impermeable or not, while
their locations can be regular or random.  From the numerical point of
view, sphere packs are particularly attractive as the distance from
any point to the surface of a sphere can be easily calculated by
knowing only the location and radius of the sphere.  When the number
of spheres is large (say, above few thousands), the computation of the
distance to the boundary of the pack as the minimum over all distances
to individual spheres becomes too time-consuming.  In that case, one
needs additional ``tools'' in order to rapidly locate a limited number
of spheres that are the closest to a given point (the current position
of Brownian trajectory).  Among such tools, one can mention
``coarse-grained'' distance maps \cite{Ossadnik91}, Whitney
decomposition or Voronoi cells.  We illustrate a Whitney decomposition
which is a partition of the mediu into squares/cubes such that the
size of each square/cube is proportional to the distance from that
element to the boundary of the medium (Fig. \ref{fig:DLA_Whitney}).
Once such a decomposition of the domain is constructed (prior to
launching Monte Carlo simulations), the distance estimation is a very
rapid: one determines the square/cube to which a given point belongs,
and takes the size of this element as an estimate for the distance.
In practice, one uses dyadic division into squares/cubes, while the
decomposition is stopped when a chosen finest resolution is reached.
The adaptive character of Whitney decompositions or other
``coarse-grained'' distance maps allows one to rapidly compute
distances for sphere packs with millions of spheres.  The related FRW
algorithm can be used for simulating growth phenomena such as
diffusion-limited aggregation \cite{Witten81,Ossadnik91,Halsey00},
diffusion-reaction processes, search problems, etc.

\section{Applications, extensions and perspectives}
\label{sec:extensions}

\subsection{Diffusion-weighted magnetic resonance imaging}

Diffusion-weighted MRI is a widespread experimental technique in which
the random trajectories of diffusing nuclei (e.g., water molecules)
are encoded by applying an inhomogeneous magnetic field
\cite{Grebenkov07,Callaghan,Price}.  When the nuclei diffuse inside a
heterogeneous medium, the statistics of random trajectories is
affected by the presence of walls or obstacles.  Although these
microscopic restrictions are not visible at the spatial resolution of
DMRI, these geometrical features are statistically aggregated into the
macroscopic signal.  Measuring the signal at different diffusion times
and magnetic field gradients, one aims to infer the morphological
structure of a sample and to characterize the dynamics of a system.
The non-invasive character of DMRI made this technique the gold
standard in material sciences, neurosciences and medicine, with
numerous applications to mineral samples (e.g., sedimentary rocks in
oil industry; soils in agriculture; concrete, cements and gypsum in
building industry) and biological samples (e.g., brain, lungs, bones).

Until nowadays, modeling DMRI experiments was mainly restricted to
simple structures \cite{Balinov93,Coy94,Duh01,Valckenborg02}, while
truly multiscale three-dimensional porous media such as concretes or
sedimentary rocks remained out of reach due to the lack of efficient
numerical techniques.  The efficiency and flexibility of FRW
algorithms make them promising tools for modeling DMRI.  An
approximate implementation of the gradient encoding into FRW
algorithms was recently proposed that opens new opportunities in the
study of mineral and biological samples
\cite{Grebenkov11,Grebenkov13b}.

\subsection{Continuous-Time Random Walks}

Many physical, chemical and biological transport processes exhibit
anomalous features, e.g., the subdiffusive scaling of the mean square
displacement \cite{Bouchaud90,Metzler00,Havlin02}.  Examples are the
motion of organelles, vesicles or tracers in living cells
\cite{Saxton97,Golding06,Metzler09,Szymanski09,Jeon11,Bertseva12},
animals searching for food \cite{Viswanathan96}, contaminant or
pollution spreading \cite{Kirchner00,Scher02}, etc.  For instance,
overcrowding in living cells and colloidal gels, or deep wells in the
interaction energy landscape \cite{Bouchaud90,Korb11}, may result in a
heavy-tailed distribution of waiting times between jumps.  Such long
stalling periods would yield subdiffusive dynamics that can be
described by Continuous-Time Random Walk (CTRW) with diverging mean
waiting time but finite-variance spatial displacements.  In that
frame, the diffusion equation (\ref{eq:diffusion}) has to be replaced
by fractional diffusion equation, in which the history of trapping or
caging is incorporated through the fractional Riemann-Liouville
derivative in time.  Since the spatial dynamics is still governed by
the Laplace operator, only the temporal dependence in the spectral
representation (\ref{eq:spectral}) has to be modified as
\begin{equation}
G_t(\r_0,\r) = \sum\limits_{m=1}^\infty E_{\alpha}(-D\lambda_m t) u_m(\r_0) u_m^*(\r) ,
\end{equation}
where $E_{\alpha}(z)$ is the Mittag-Leffler function \cite{Haubold11}.
Using this extension, one can recompute the exit time and position
distributions in order to simulate restricted subdiffusion in
multiscale or complex structures
\cite{Yuste07,Grebenkov10a,Grebenkov10b}.

\subsection{Intermittent processes}

Another important extension of FRW is related to diffusive processes
with two successively alternating phases (e.g., active and passive
transport of vesicles in living cells
\cite{Caspi00,Arcizet08,Brangwynne09,Kenwright12}).  These so-called
intermittent processes have been intensively studied during the last
decade.  In particular, the alternation of phases was shown to
facilitate search processes and speed up biochemical kinetics (see the
review \cite{Benichou11} and references therein).  For instance, in
the case of surface-mediated diffusion in spherical domains, the mean
time for finding a target was shown to be minimal when the phases of
bulk and surface diffusion alternate at certain rate
\cite{Benichou10,Benichou11b,Calandre12,Rupprecht12}.  At the same
time, the question of optimality for the mean search time remains open
for porous media or irregularly-shaped domains such as catalysts or
biological structures.  The FRW algorithm can be adapted to simulate
such intermittent diffusive processes.

\subsection{Conclusions and Perspectives}

In this chapter, we described the basic principles and several
applications of fast random walk algorithms.  The algorithm relies on
adaptive splitting of the random trajectory inside a complex medium
into independent ``elementary blocks'' in simple jump domains for
which simulations can be performed much more efficiently.  In other
words, the algorithm eliminates the geometrical complexity of the
system and reduces the original problem to the study of diffusion
inside a disk, a sphere or a rectangle, for which many mathematical
tools (PDE, spectral methods, etc.)  are particularly efficient
\cite{Crank,Carslaw}.  This ``trick'' drastically speeds up Monte
Carlo simulations since at each step the largest possible exploration
is performed.  In practice, the computation is reduced to estimating
the distance from any interior point to the boundary that can be
solved in different ways depending on the studied geometry.  The
adaptive character of FRW techniques makes them well suited for
simulating diffusion-reaction processes in complex, multi-scale,
disordered, heterogeneous or irregularly-shaped media.  The FRW
algorithms can be applied for calculating the harmonic measures and
their extensions, the first passage and exit time distributions,
search times, reaction rates \cite{Agmon84,Weiss86}, residence times
\cite{Grebenkov07a}, etc.  Many growth processes such as
diffusion-limited aggregation and related models can be efficiently
realized by using these techniques.  We also illustrated one
application of these algorithms for computing DMRI signals and
discussed extensions for modeling restricted anomalous diffusion and
intermittent processes.

\end{document}